\documentclass[preprint,showpacs,preprintnumbers,amsmath,amssymb]{revtex4}

\usepackage{graphicx}% Include figure files
\usepackage{dcolumn}% Align table columns on decimal point
\usepackage{bm}% bold math

\begin{document}

\preprint{}

\title{Micrometric particles 2D self-assembly during drying of liquid film}

\author{ F.Lallet$^1$ and N.Olivi-Tran$^{1,2}$} 
\affiliation{$^1$Laboratoire de Sciences des Proc\'ed\'es C\'eramiques et Traitements de Surface, UMR-CNRS 6638, Ecole Nationale Sup\'erieure de C\'eramiques Industrielles, 47 avenue Albert Thomas, 87065 Limoges cedex,
France \\ $^2$Laboratoire de Physique de la Mati\`ere Condens\'ee et Nanostructures, UMR-CNRS 5586, Universit\'e Claude Bernard Lyon I, Domaine Scientifique de la Doua, 69622 Villeurbanne cedex, France}

\date{today}

\begin{abstract}
We computed the self-organisation process of a monodisperse collection 
of spherical micrometric particles trapped in a two-dimensional (2D) thin
 liquid film isothermally dried on a chemically inert substrate. The substrate
 is either flat or indented to create linear stripes on its surface. 
The numerical results are illustrated and discussed in the light of 
 experimental ones obtained from the drying of diamond particles water based 
suspension ($d_{50} = 10 \mu m$) on a glass substrate. The drying of the
 suspension on a flat substrate leads to the formation of linear patterns
 and small clusters of micrometric particles distributed over the whole
 surface of the substrate, whereas the drying of the suspension on a 
indented substrate leads to the aggregation of the particles along one side of the stripe which has a higher roughness than the other side of the stripe. This is an easy experimental way to obtain colloidal selforganized patterns. 
\end{abstract}

\vfill
\maketitle
\section{Introduction}

The fabrication of self-organized 3D crystals like opals \cite{Kumaraswamy} \cite{Fudouzi} \cite{Brozell}, 2D dense monolayers or patterns \cite{Ray} \cite{Gunther} \cite{Yarin} \cite{Bigioni} of micrometric and more recently nanometric particles is an intense field of scientific researches.

In this paper, we focus on the 2D patterns formed after the horizontal drying
 of a thin liquid film containing micrometric particles. Generally speaking, 
the self-organisation of particles trapped in a 2D thin liquid film is induced
 during the last stage of drying through the action of lateral capillary 
forces. As reported by Paunov \cite{Paunov}, those forces are analogous
 to electrostatic interactions and "[...] result from the overlapping of the 
liquid surface deformations around the particles". The liquid surface
 deformations appear when the particles are no more covered but surrounded
 by the liquid film at the last stage of drying. The overviews of Kralchevsky
 \cite{Kralchevsky1} \cite{Kralchevsky2} and Nagayama \cite{Nagayama} 
point out that these interactions are fundamental to understand the
 self-patterning mechanisms of particles trapped in a 2D thin film dried
 on a substrate in most cases. However, the self-patterning processes of
 particles submitted to lateral capillary forces can be balanced by
 the interactions between the particles and the substrate. For example, 
the lateral capillary forces can be correlated to particle - substrate electrostatic interactions. 
Indeed, Aizenberg \cite{Aizenberg} and Huwiler \cite{Huwiler} succeeded 
to create original 2D patterns of polystyrene microspheres and silica 
nanoparticles  through template recognition processes using chemically
 pre-patterned substrates (lithographically modified substrates). Moreover,
 Thill \cite{Thill} demonstrated that the adhesion force between
 $\alpha$-$Al_{2}O_{3}$ particles and a mica substrate can overcome the effects
 of lateral capillary interactions and stick the particles on the surface of the substrate.
 The adhesion force ($L$) which appears when a particle is deposited on the
 surface of a substrate might induce friction forces
 ($F$) through the classical Amontons' law \cite{Ecke}, $F=\mu L$,
 where $\mu$ is a friction coefficient. In particular,
 Hu \cite{Hu} demonstrated that low friction forces between silica 
nanoparticles ($d_{50}=25nm$) moving on a silicon substrate are 
a key point to reach a well distributed self-ordered 2D particles close-packings on the surface of the substrate at the end of drying.

However, despite many experimental results on particles self-organisation,
 few theoretical models have been reported in the literature to describe
 dynamically the physical phenomena occurring during the drying of a thin 
liquid film containing micrometric particles. Popov described the drying stage 
of a colloidal suspension droplet on a substrate through a complete analytical
 model \cite{Popov}, Reyes studied the geometrical properties of polymer colloidal particles monolayers synthesized from the drying of a colloidal suspension by a Monte Carlo approach \cite{Reyes} and J\'arai-Szab\'o proposed a spring-block stick-slip model to reveal the self-ordering dynamics of polystyrene nanoparticles on a glass substrate \cite{Jarai-Szabo}. But, to the best of our knowledge, there is no simple dynamic model able to describe the physical events occurring during the drying of a thin liquid film containing nanometric or micrometric particles. 
 
Thus, the aim of our study is to propose a numerical model, which allows one to simulate the physical events occurring during the drying of a thin liquid film containing micrometric particles, in order to characterize their distribution at the end of drying. The thin liquid film is horizontally dried on a substrate which is either flat or indented to create linear stripes on its surface. The stripes are made in such a way that a rough area is also created on the border of one side of each stripe. By this way, we will point out the critical effects of lateral capillary forces and particles-substrate friction forces on the self-organization process of particles with regard to a rough area at the top of the substrate and a stripe in the substrate. Therefore, we have developed a molecular dynamic algorithm which allows one to calculate the trajectories of the particles by solving the Newton's equation of motion across time during drying until the thin film is fully dried. Moreover, we present a simple experimental procedure to illustrate the drying of a thin film on a substrate either flat or indented. 

The first section of the paper deals with the experimental procedure while the second one describes the physical model involves in molecular dynamic calculations. The third section presents the numerical results which are discussed in the light of experimental data.

\section{Experimental approach}

\subsection{Synthesis of micrometric particles patterns}
The particles are faceted diamond monocrystals with an average diameter of 10 $\mu$m. The experimental suspension is made from 2 ml of a commercial water based diamond suspension Buehler Metadi\textsuperscript{\textregistered} diluted  in 40 ml of distilled water. The number of particles per unit volume in this suspension is estimated to $10^7$ particles for $1 cm^3$ of liquid. We used amorphous glass substrates ($1\times 1\times 0.1 cm^{3}$) with an optical quality surface. The substrate is either fully flat or locally indented with a diamond pen to create linear stripes on its surface. The action of the diamond pen at the surface of the substrate leads to the formation of a rough  surface on one side of the stripe. The figure 1 gives a schematic description of the substrate next to the stripe. Afterwards, the substrates are washed in soapy water before being cleaned under ultrasonic vibrations in water/acetone and water/ethanol bathes.

The glass substrate, either flat or indented, is deposited in a container filled with 600 $\mu$L of the suspension. Then this system is deposited in a drying oven which temperature is constant and equal to 60\textsuperscript{o}C. The container is removed from the drying oven when the suspension is fully dried.

\subsection{Optical microscopy analysis of the patterns} 
We used an optical microscope in transmission mode to observe the micrometric particles distribution on the surface of the substrate after drying.
Figure 2 is an image of a micrometric particles collection obtained after drying of the suspension on a flat substrate. One can see that the particles are distributed over the whole surface of the substrate. This picture clearly shows particles which are either isolated or aggregated in small clusters or interconnected into linear patterns. The figure 3 is also a snapshot of a micrometric particles collection obtained after drying but in the case where the substrate is indented; the width of the stripe is about 3 times the average diameter of the particles. This picture shows that there is a dramatic effect of the rough edge of the stripe on the distribution of the particles. We do not observe small clusters or linear patterns any more but some single particles distributed around a dense band of interconnected particles located along one side of the stripe. The width of this band is estimated to be equal to 4-5 times the average diameter of the particles.

On  one hand,  figure 2 allows us to think that at the end of the drying process (when the thickness of the film is lower than the average diameter of the particles), the thin liquid film breaks into several non-spherical drops more or less interconnected \cite{Thouy}.  Afterwards, the particles which are located in a liquid drop are submitted to a local confinement due to lateral capillary forces. They aggregate with their nearest neighbors which leads to the formation of small clusters or linear patterns when the surface of the substrate is flat.

On the other hand, one can see a dense band of particles located along the stripe in the figure 3. One can see that the aggregation of the particles is not symmetric with regard to the stripe. Thus, we can not explain this tendency only through the action of lateral capillary forces as those interactions are isotropic. Therefore,  friction forces, induced by the higher roughness on one side of the stripe may  be responsible for the particles preferential location. Moreover, there are not any particles inside the stripe. Indeed, as the thin liquid film is dried on the top of the substrate but not inside the stripe, the particles located next to the stripe are submitted to stronger capillary interactions than the particles located inside the stripe. Therefore, the particles do not stay inside the stripe but are preferentially trapped on the rough band next to the stripe where the action of friction forces is enhanced by the action of lateral capillary forces.    

\section{Numerical simulation}
\subsection{Model of thin liquid film and substrates}

The previous section introduced the dramatic effect of the stripes on the distribution of the particles at the surface of the substrate after drying. We propose a physical model to understand this difference.
We used molecular dynamic simulations at constant temperature (333K - 60\textsuperscript{o}C) to solve  Newton's equation of motion in order to simulate the trajectories of the particles as a function of time until the thin liquid film is fully dried. The integration time step is $\Delta t=10\textsuperscript{-10}s$.

A numerical thin liquid film is simulated with a collection of $N=200$ spherical particles of equal diameter $d=10 \mu m$ and mass $m$ randomly distributed in a box of length $50d$ in the $x$ and $y$ axis and height $h=d$ on the $z$ axis at the beginning of the calculation process. We suppose that there is no displacement in the $z$ direction, the particles are trapped in a 2D thin liquid film and move at the surface of the substrate in the $x$ and $y$ directions. One can note that spherical particles can either slide or roll or both at the surface of the substrate. However, it is clear that the faceted experimental particles slide on the surface of the substrate. Therefore we assume that the virtual spherical particles slide without rolling at the surface of the substrate, which induces particles-substrate friction forces. In order to simulate the effect of drying, the height $h$ of the air-liquid interface is decreased linearly at each step from $h=d$ at the beginning to $h=0$ at the end of the calculation process. Therefore, the air-liquid and liquid-solid interfaces of the film are characterized by the ($x$,$y$,$h$) and ($x$,$y$,$0$) planes respectively. Whatever the kind of substrate which is considered, we applied periodic boundary conditions in the $x$ and $y$ directions.

The thin liquid film is either dried on a fully flat substrate or on a substrate indented. In that case we introduce the effect of the stripe. The defect (i.e. the stripe), located at the center of the substrate,  is represented by a rectangular band of width $3d$ on the $x$ axis, length $50d$ on the $y$ axis and depth $3d$ on the $z$ axis. The particles located in this defect are allowed to move in the three directions of space. As in experiments, the enhanced roughness is only available along the left side of the stripe. Therefore, we propose to simulate the enhanced roughness on a border of the stripe by a friction force induced by a higher local roughness with regard to the other parts of the substrate. In order to simulate this roughening effect on one side of the stripe, we introduce an area of width $5d$ on the $x$ axis and length $50d$ on the $y$ axis, located on the border of the left side of the stripe at the top of the substrate, with a higher roughness than on the other parts of the substrate. 

Whatever the kind of substrate, the computation process is stopped when $h=0$ which implies that the particles located in a stripe are still surrounded by the liquid medium. We do not simulate the drying of the stripe because this stage has no effects on the particles distribution characteristics at the surface of the substrate.

\subsection{Computation of the forces involved in the drying process}
As the particles are very small, we suppose that the effect of  weight on their movement can be neglected in the whole calculation process.

The particles which are in thermal equilibrium with the liquid medium are submitted to Brownian motion which is linked to a viscous drag force according to the fluctuation-dissipation theorem. Those forces are valid for particles moving at the surface of a substrate or inside a stripe. The Brownian motion is modelled through a Gaussian random force, uncorrelated neither in space nor in time, whose mean with regard to time is expressed with the following expression:
\begin{equation}
\langle f_{i}\textsuperscript{b}(t),f_{i}\textsuperscript{b}(t') \rangle=2k_B T\xi \delta(t-t')
\end{equation}
where $k_B$ is the Boltzmann's constant, $T$ the absolute temperature and $\xi$ ($kg.s\textsuperscript{-1}$) a friction coefficient which is related to the physical parameters of the system, $\xi=3\pi d\eta$, with $\eta$ ($kg.m\textsuperscript{-1}.s\textsuperscript{-1}$) the dynamic viscosity of the liquid medium. The last term, $\delta$, is the Dirac's distribution function. We suppose that the liquid is pure water whose dynamic viscosity is $\eta=0.5\times 10\textsuperscript{3}kg.m\textsuperscript{-1}.s\textsuperscript{-1}$ at 333K.
In order to compute the Brownian motion, the Gaussian random force must be expressed through the formalism of Wiener:
\begin{equation}
f_{i}\textsuperscript{b}(t)\Delta t=\sqrt{6\pi k_BTd\eta}\Delta W(t)
\end{equation} 
where $\Delta W(t)$ is computed by a Gaussian random number as $\langle \Delta W(t) \rangle=0$ and $\langle \Delta W(t)\Delta W(t') \rangle=\Delta t$.
The Brownian motion involves a viscous drag force which can be formalized as follows:
\begin{equation}
f_{i}\textsuperscript{s}(t)=-\xi v_i(t)=-3\pi d\eta v_i(t)
\end{equation}
with $v_i(t)$ is the velocity of the particle $i$ at time $t$.
The drying process induces the decreasing of the thickness of the thin liquid film and the particles located on a fully flat substrate or outside the stripe of an indented substrate are not immersed but surrounded by the liquid medium. Consequently the Brownian motion and the viscous drag forces decrease (the contact area between each particle and the liquid medium is lowered). We suppose, for the sake of simplicity, that the intensities of those forces decrease linearly at each step of the calculation process which implies that $d$ is replaced by $h$ in equations (2) and (3).

The particles which are trapped in a two dimensional thin film of liquid are subjected to lateral capillary forces (those capillary forces are not relevant for particles located in the depth of  a stripe because they are surrounded by the liquid medium during the whole calculation process).
The lateral capillary forces are computed with the following expression \cite{Aizenberg}:
\begin{equation}
f_{i}\textsuperscript{c}(r)=-\pi\gamma r_c^{2}sin(\phi_c)^{2}\frac{d}{(r-d)^{2}}
\end{equation}
where $\gamma$ ($N.m\textsuperscript{-1}$) is the surface tension of water $\gamma=73\times 10\textsuperscript{-3}N.m\textsuperscript{-1}$ at 293K, $r_c$ the radius of the liquid-solid contact line and $\phi_c$ the mean slope angle at the meniscus at the contact line. Figure 4 presents a schematic description of the physical parameters involved in the calculation of the lateral capillary interactions. The analytical expressions of $r_c$ and $\phi_c$ are \cite{Aizenberg}:
\begin{eqnarray}
r_c & = & (h(2R-h))^{\frac{1}{2}} \\
\phi_c & = & arcsin(\frac{r_c}{R})-\alpha
\end{eqnarray}
where $R$ is the radius of the particle and $\alpha$ the wetting angle at the three contact line. We suppose that the particles are wetted by the liquid medium and we choose $\alpha=0.7rad$ ($~40\textsuperscript{o}$).
The lateral capillary forces are spatially limited in the calculation process. Indeed, it is clear that the overlapping of the deformations induced by the presence of a particle is efficient only for its next neighbors. Therefore we propose to compute the lateral capillary forces with the following conditions:
\begin{eqnarray}
\textrm{if}\qquad r & > & 5d\qquad \textrm{then:} \nonumber\\
f_{i}\textsuperscript{c}(r) & = & 0 \\
\textrm{if}\qquad r & \leq & 5d\qquad  \textrm{then:} \nonumber\\
\textrm{if}\qquad r & > & 1.1d\qquad  \textrm{then:} \nonumber\\
f_{i}\textsuperscript{c}(r) & = & -\pi\sigma r_c^{2}sin(\phi_c)^{2}\frac{d}{(r-d)^{2}} \\
\textrm{if}\qquad r & \leq & 1.1d\qquad \textrm{then:} \nonumber\\
f_{i}\textsuperscript{c}(r) & = & -\pi\sigma r_c^{2}sin(\phi_c)^{2}\frac{d}{(1.1d-d)^{2}}
\end{eqnarray}
Finally, we consider the friction forces induced by the roughness of the substrate. As previously reported, we modelled the faceted particles of the real system by spherical particles in the virtual system which slide at the surface of the substrate. It is clear that the face of a real particle which is in contact with the surface of the substrate is responsible for the development of particle-substrate friction forces. Thus, to be consistent with the real system, the virtual particles are submitted to friction forces. Therefore, at the end of each calculation step, the total interaction supported by a virtual particle is expressed as:
\begin{equation}
f_{i}\textsuperscript{t}(r,t)=(1-\theta)(f_{i}\textsuperscript{b}(t)+f_{i}\textsuperscript{s}(t)+f_{i}\textsuperscript{c}(r))
\end{equation}
with $\theta$ an dimensionless friction coefficient. For the sake of simplicity, we assume that if the particle moves at the surface of a fully flat substrate or out of the rough area of an indented substrate $\theta=0$, whereas if the particle is located on the rough part of an indented substrate, $\theta=0.9$. We chose a very high friction coefficient to point out clearly the effects of friction forces on the particles distribution between a flat and an indented substrate.

\section{Experimental and numerical results: analysis and discussion}
 
Figure 5 is a numerical particles distribution which corresponds to the final state of a numerical thin liquid film dried on a fully flat substrate whereas  figure 6 presents a numerical particles distribution associated to an indented substrate. One can clearly see that the calculations demonstrate a good qualitative agreement with experimental data. Indeed, similarly to figure 2, linear patterns, small clusters and even single numerical particles are observed in figure 5. Moreover, like in figure 3, a dense band of numerical particles located on the rough part of the substrate is observed in figure 6. Therefore, whatever the kind of substrate, the physical model introduced in section 3 is consistent to describe the evolution of the system until the thin liquid film is fully dried.

On the one hand, the numerical results point out that the roughness effects are enhanced by the lateral capillary forces at the end of the drying process. Indeed, as previously mentioned, the last stage of drying of a thin liquid film on a perfect substrate leads to the formation of drops more or less interconnected which are responsible for the small patterns of particles. This phenomenon is modelled in the numerical approach through the action of the lateral capillary interactions which are spatially limited. Consequently, the numerical particles can only interact with their nearest neighbors to form small clusters or linear patterns at the surface of the substrate. 

On the other hand, the numerical results demonstrate that the local roughness of the substrate and the friction forces are relevant to simulate the confinement effect close to the stripe. Whatever the case (real or virtual system), the lateral capillary interactions stick the particles with their very next neighbors. In the case of a substrate with linear stripe, we demonstrate that the confinement effect along one side of the stripe is physically induced by friction forces.   

The figure 7 shows the space correlation function of the particle distribution in real space. One may see that for distances enclosed between 1 and 4, the density of presence of particles is larger in the case of an indented substrate (bottom) than in the case of a flat substrate (top). This mean that the particles are preferentially agglomerated in some location, in our case a large number of particles are located on the left part the stripe.
Moreover, one may see that the density of presence in the bottom figure, (for abscissa  lower than 50 corresponding to an interparticle distance of 50), is the signature of presence of stripes in the $y$ direction. The fact that both distribution functions decrease after interparticle distance equal to 70 is due to the computation of the correlation functions which does not take into account periodic boundary conditions.

\section{Conclusion}

We showed that the effect of roughness and thus of friction is enhanced by the action of lateral capillary forces.
By imprinting linear defects, where the roughness of the substrate is larger, we obtained after drying of the thin liquid film linear stripes of micrometric particles. We demonstrated that a rough surface at the top of a substrate is a better way than stripes to confine the particles.
Experiments and molecular dynamic simulations are in good agreement. This experimental method is an easy way to obtain correlated patterns of faceted micrometric particles.

Acknowledgments
The authors acknowledge T.Albaret for helping experimental images format conversion. 
\pagebreak

\begin{figure}
\includegraphics[width=14cm]{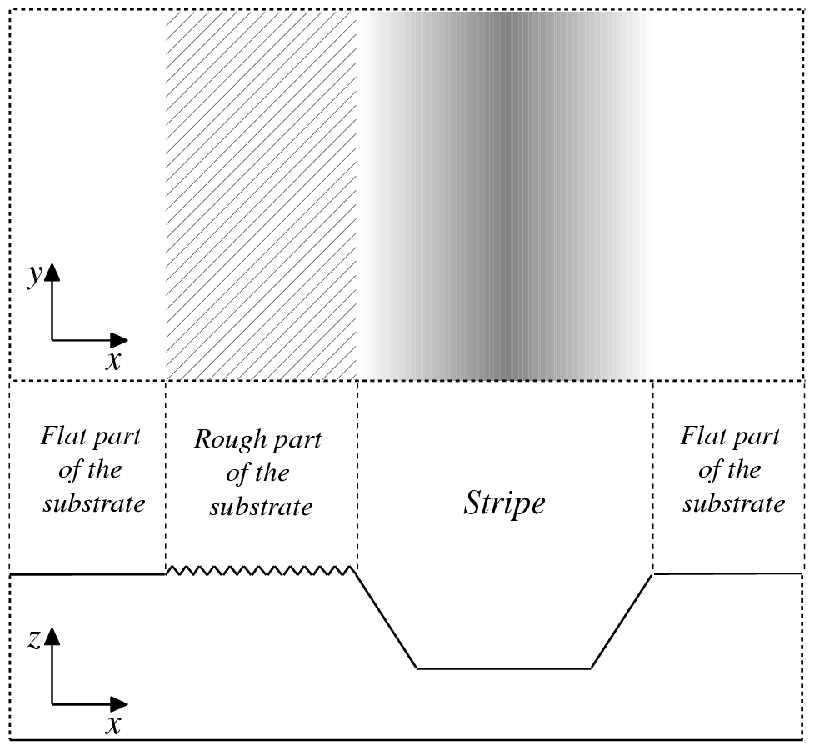}
\caption{Schematic representation of the stripe and the rough part of the substrate indented from a top view (top) and section view (bottom).}
\end{figure}

\begin{figure}
\includegraphics[width=14cm]{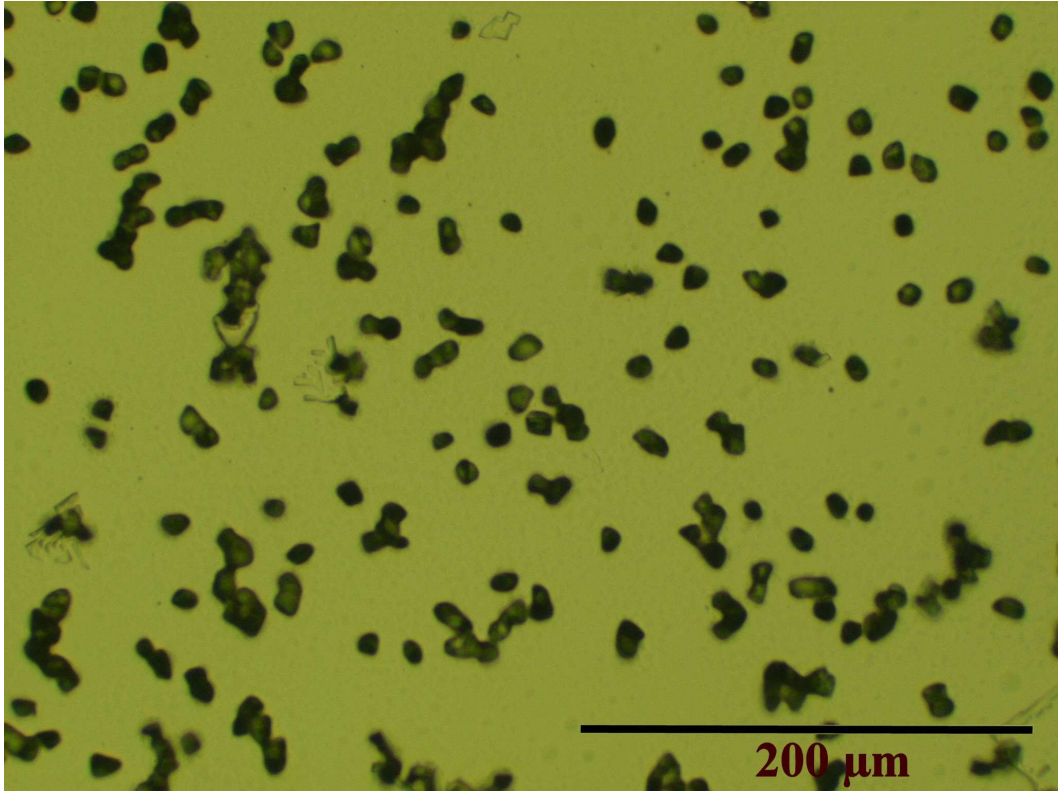}
\caption{Patterns of particles observed by optical microscopy in transmission mode after drying on a flat glass substrate.}
\end{figure}
 
\begin{figure}
\includegraphics[width=14cm]{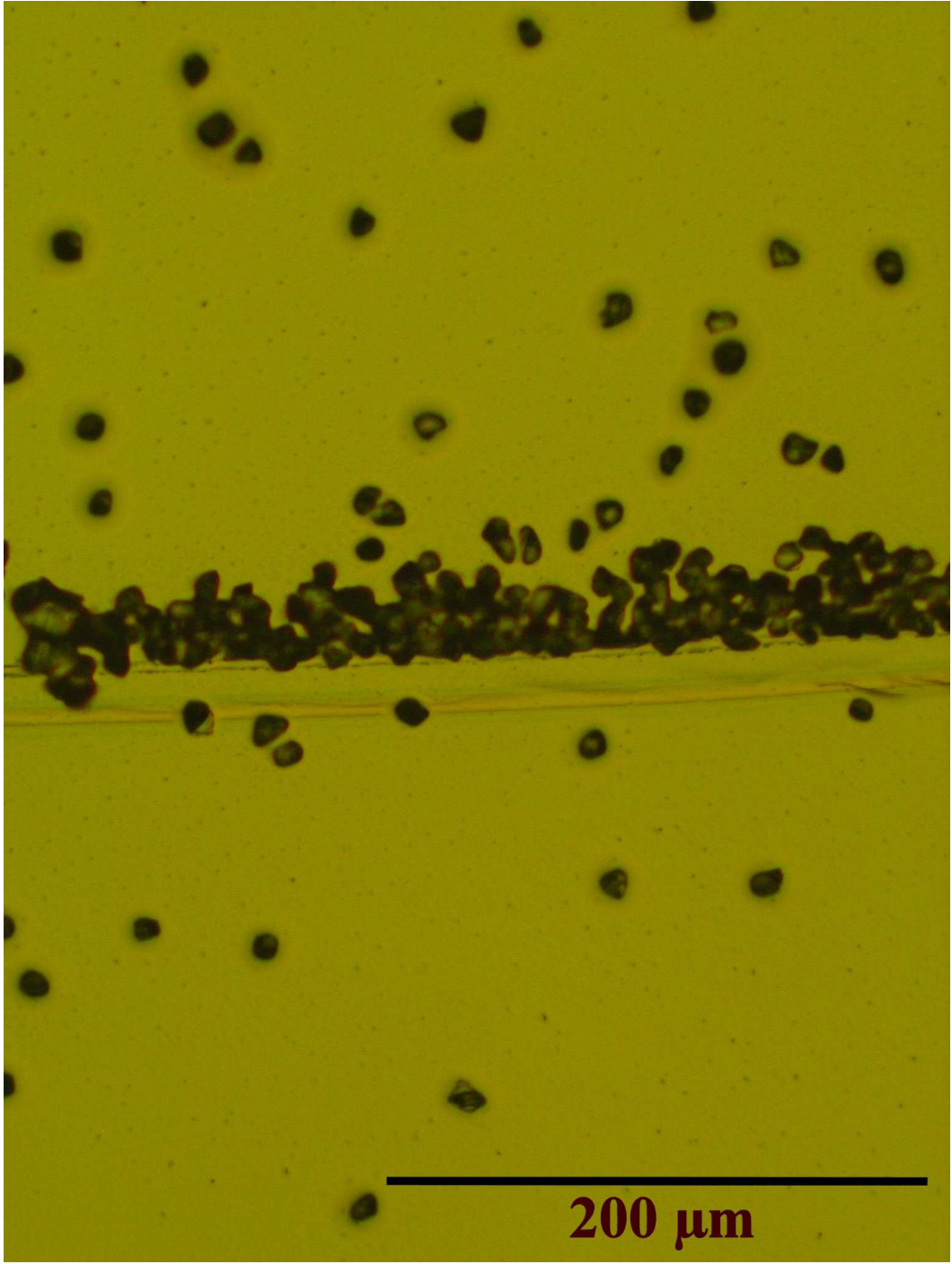}
\caption{Patterns of particles observed by optical microscopy in transmission mode after drying on an indented glass substrate. Experimental evidence of a confinement effect along one side of the stripe.}
\end{figure}

\begin{figure}
\includegraphics[width=14cm]{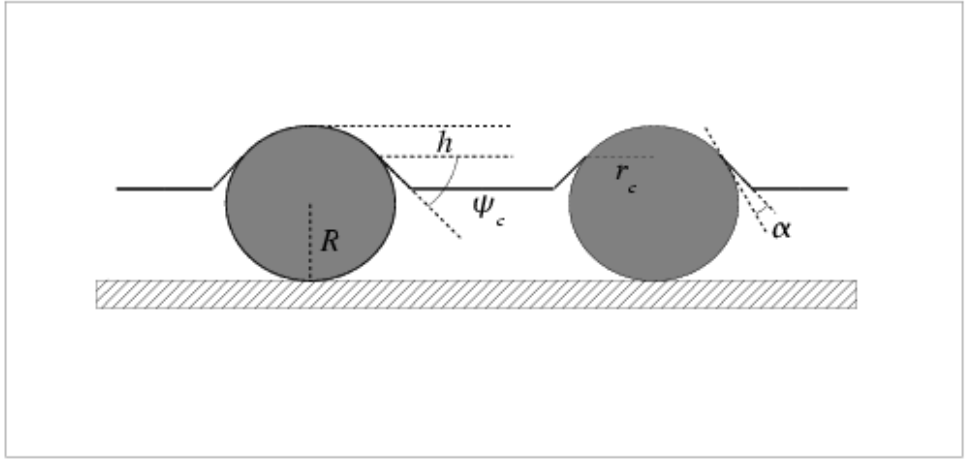}
\caption{Schematic description of the physical parameters involve in the computation of the lateral capillary interactions for two spherical particles of equal diameter trapped in a thin liquid film.}
\end{figure}

\begin{figure}
\includegraphics[width=14cm]{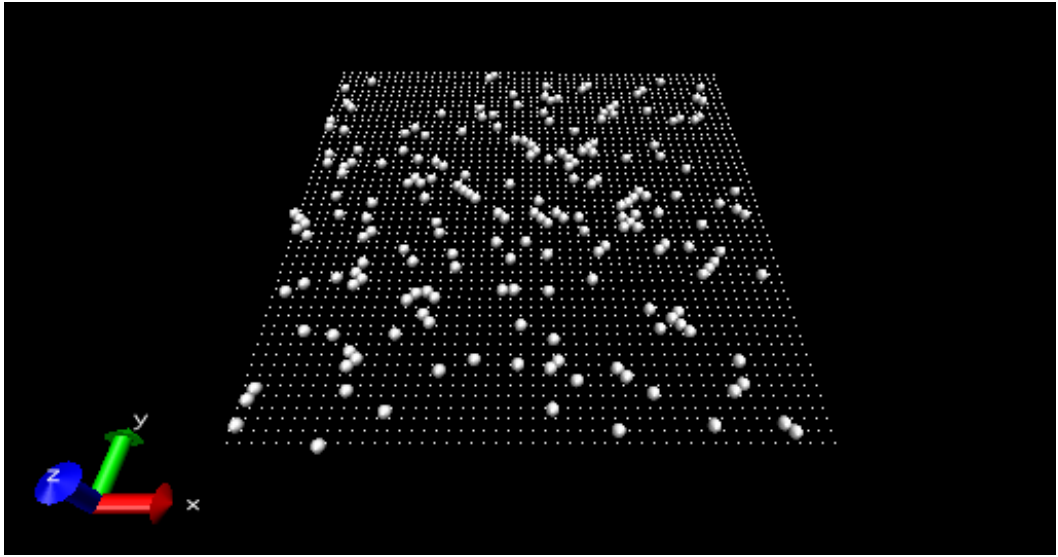}
\caption{Numerical patterns of particles after drying on an flat substrate.}
\end{figure}

\begin{figure}
\includegraphics[width=14cm]{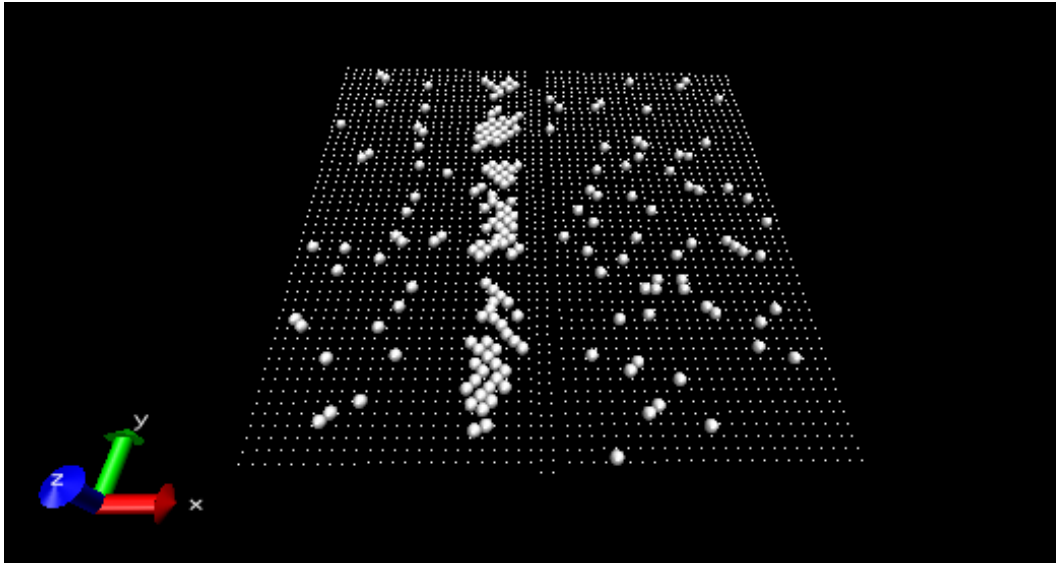}
\caption{Numerical patterns of particles after drying on an substrate indented.}
\end{figure}

\pagebreak

\begin{figure}
\includegraphics[width=10cm]{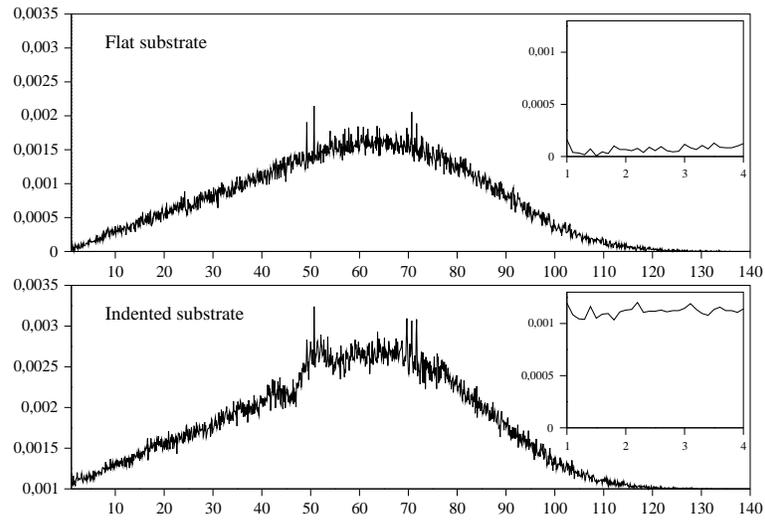}
\caption{Distribution function of pairs for the numerical distribution of particles on a flat substrate (top) and on an indented substrate (bottom).}
\end{figure}

\end{document}